\documentclass[aps,twocolumn,epsfig,prl,superscriptaddress]{revtex4}

\usepackage{amsmath}
\usepackage{graphicx}

\begin{document}

\title{Hamiltonian decomposition for bulk and surface states}

\author{Ken-ichi Sasaki}
\email[Email address: ]{sasaki@hiroshima-u.ac.jp}
\affiliation{Department of Quantum Matter, Graduate School of Advanced
Sciences of Matter (AdSM), Hiroshima University, Higashi-Hiroshima
739-8530, Japan} 

\author{Yuji Shimomura}
\affiliation{Department of Quantum Matter, Graduate School of Advanced
Sciences of Matter (AdSM), Hiroshima University, Higashi-Hiroshima
739-8530, Japan} 

\author{Yositake Takane}
\affiliation{Department of Quantum Matter, Graduate School of Advanced
Sciences of Matter (AdSM), Hiroshima University, Higashi-Hiroshima
739-8530, Japan} 

\author{Katsunori Wakabayashi}
\affiliation{Department of Quantum Matter, Graduate School of Advanced
Sciences of Matter (AdSM), Hiroshima University, Higashi-Hiroshima
739-8530, Japan} 
\affiliation{PRESTO, Japan Science and Technology Agency (JST),
Kawaguchi 332-0012, Japan}

\author{}
\affiliation{}

\date{\today}
 
\begin{abstract}
 We demonstrate that a tight-binding Hamiltonian 
 with nearest- and next-nearest-neighbor hopping integrals
 can be decomposed into bulk and boundary parts
 in a general lattice system.
 The Hamiltonian decomposition reveals that 
 next nearest-neighbor hopping causes sizable changes
 in the energy spectrum of surface states 
 even if the correction 
 to the energy spectrum of bulk states is negligible.
 By applying the Hamiltonian decomposition to 
 edge states in graphene systems,
 we show that the next nearest-neighbor hopping
 stabilizes the edge states.
\end{abstract}

\maketitle

The energy band structure is of central importance
in understanding the electronic properties of material.
A tight-binding (TB) model is a versatile approach
to study the electronic, 
magnetic and transport properties of solid
since TB model describes the qualitative features 
of the energy band structure.~\cite{kittel04}
The long-range hopping terms such as 
next nearest-neighbor (nnn) hopping are often added to 
the TB model with nearest-neighbor (nn) hopping
to improve the energy band structure.
In many cases, the nnn correction 
to the energy band structure
is not a matter of particular importance.
In this Letter, we show that
the nnn hopping can change appreciably 
the energy spectrum of surface states which appear
near the boundary of a system
even when the correction 
to the energy spectrum of bulk states is negligible.
We explain this by decomposing the nnn TB Hamiltonian
into two parts; bulk and boundary parts.
This Hamiltonian decomposition is 
essential to understanding 
the stability of surface states.

We use graphene systems
to demonstrate the Hamiltonian decomposition 
for the following reasons.
(1) Graphene is known to have both bulk and surface states.
The bulk states exhibit 
a ``relativistic'' energy band structure 
called the Dirac cone,~\cite{wallace47,slonczewski58} and 
the surface states called the edge states
appear near the zigzag edge.~\cite{fujita96}
(2) The edge states have been observed 
by several experimental
groups,~\cite{klusek00,niimi05,kobayashi05,sugawara06}
and a theoretical understanding of the experimental results
is called for.
In fact, the nnn hopping is important 
to explain the experimental results.~\cite{sasaki06apl}
(3) The edge states have
a large density of states (DOS) near the Fermi energy, which
is responsible for the Fermi instabilities.
Since the DOS depends on the the energy spectrum or 
bandwidth of the edge states, the nnn hopping
is important for the appearance of many-body effects
of the edge states.~\cite{sasaki06super,sasaki08jpsj}
Thus, graphene is a good testing system for 
the Hamiltonian decomposition not only from a theoretical
but also from an experimental point of view.

We study the nn (nnn) TB Hamiltonian,
$H_{\rm nn}$ ($H_{\rm nnn}$), which is defined as
\begin{align}
 \begin{split}
  & \left( \frac{H_{\rm nn}}{-\gamma_0} \right)
  = \sum_{i,j\in {\rm all}}
  c_i^\dagger [{\cal H}_{\rm nn}]_{ij} c_j, \\
  & \left( \frac{H_{\rm nnn}}{-\gamma_n} \right)
  = \sum_{i,j\in {\rm all}}
  c_i^\dagger [{\cal H}_{\rm nnn}]_{ij} c_j,
  \label{eq:nn}
 \end{split}
\end{align}
where $c_i$ ($c_i^\dagger$) is the annihilation (creation) operator
of an electron at $i$-th site,
$\gamma_0$ ($\gamma_n$) is the nn (nnn) hopping integral, and
the matrix element $[{\cal H}_{\rm nn}]_{ij}$ 
($[{\cal H}_{\rm nnn}]_{ij}$) is 1 
when $i$-th site is a (next) nn site of $j$-th site 
and is zero otherwise.
In the following, 
we will show that $H_{\rm nnn}$ can be decomposed 
into bulk and boundary (edge) parts for a graphene 
with zigzag edge.

\begin{figure}[htbp]
 \begin{center}
  \includegraphics[scale=0.4]{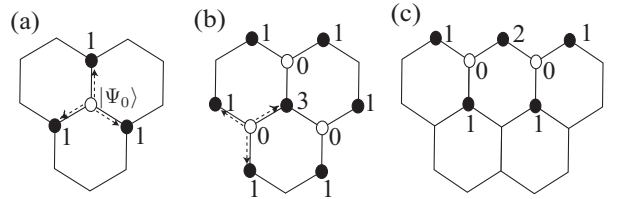}
 \end{center}
 \caption{
 (a) An electron at the central site, 
 $|\Psi_0 \rangle=c_0^\dagger|0\rangle$, 
 is transfered to three nn sites by $H_{\rm nn}/(-\gamma_0)$.
 The resultant state is 
 $|\Psi'_0 \rangle=\sum_{i\in {\rm all}} [{\cal H}_{\rm
 nn}]_{i0}c_i^\dagger|0\rangle$.
 (b) 
 ${\cal H}_{\rm nn}^2$ transfers the electron to the nnn sites. 
 At the same time, ${\cal H}_{\rm nn}^2$ returns the electron 
 to the original site.
 The matrix element of $[{\cal H}_{\rm nn}^2]_{ij}$ 
 that returns the electron to the original site 
 is given by 3 because there are three nn sites around the central site.
 (c) 
 The matrix element for the return process
 $[{\cal H}_{\rm nn}^2]_{ii}$ 
 depends on whether the site is a bulk site 
 ($[{\cal H}_{\rm nn}^2]_{ii}=3$)
 or a zigzag edge site ($[{\cal H}_{\rm nn}^2]_{jj}=2$).
 } 
 \label{fig:nnn}
\end{figure}

Suppose that we put an electron on the central site
denoted by the empty circle in Fig.~\ref{fig:nnn}(a).
The initial state is labeled as
$|\Psi_0 \rangle=c_0^\dagger|0\rangle$.
We operate on $|\Psi_0 \rangle$ with $H_{\rm nn}/(-\gamma_0)$,
then ${\cal H}_{\rm nn}$ transfers the electron 
to three nn sites denoted by the solid circles
in Fig.~\ref{fig:nnn}(a). 
This state is written as
$|\Psi'_0 \rangle=\sum_{i\in {\rm all}} 
[{\cal H}_{\rm nn}]_{i0}c_i^\dagger|0\rangle$.
The numbers associated with the solid circles in Fig.~\ref{fig:nnn}(a)
indicate the matrix element of ${\cal H}_{\rm nn}$.
The successive operation of $H_{\rm nn}/(-\gamma_0)$
on $|\Psi'_0\rangle$ gives
\begin{align}
 |\Psi''_0 \rangle=\sum_{i,j\in {\rm all}} 
 [{\cal H}_{\rm nn}]_{ji}[{\cal H}_{\rm nn}]_{i0}c_j^\dagger|0\rangle.
 \label{eq:psi''}
\end{align}
Starting from the initial site, the electron
reaches the nnn sites as shown in Fig.~\ref{fig:nnn}(b).
Thus, the two successive nn hopping processes
relate to the nnn hopping process.
This indicates that ${\cal H}_{\rm nn}^2$ 
includes ${\cal H}_{\rm nnn}$.
However ${\cal H}_{\rm nn}^2$ and ${\cal H}_{\rm nnn}$ are not identical
because in ${\cal H}_{\rm nn}^2$ there is a diagonal matrix element
that returns the electron to the original site, that is,
$|\Psi''_0 \rangle$ of Eq.~(\ref{eq:psi''}) contains the term with
$j=0$.
Since there are three nn sites around the original site,
the amplitude of this return process is 3
as shown in Fig.~\ref{fig:nnn}(b).
${\cal H}_{\rm nn}^2$ and ${\cal H}_{\rm nnn}$ 
become identical if we subtract 
the corresponding diagonal matrix element from ${\cal H}_{\rm nn}^2$.
This matrix is proportional to the unit matrix, 
$[I]_{ij}=\delta_{ij}$.

For a periodic system, 
since the number of bonds of every site is three, 
we have
${\cal H}_{\rm nnn} = {\cal H}^2_{\rm nn} - 3 I$.
Putting this into Eq.~(\ref{eq:nn}),
we see that $H_{\rm nnn}$ can be rewritten as
\begin{align}
 H_{\rm nnn} 
 = -\gamma_n \sum_{i,j\in {\rm all}}
  c_i^\dagger [{\cal H}^2_{\rm nn} - 3 I]_{ij} c_j.
 \label{eq:Hnnn}
\end{align}
The matrix ${\cal H}_{\rm nn}$ can be diagonalized 
by a unitary matrix as
$[U {\cal H}_{\rm nn} U^\dagger]_{pq}=E_p/(-\gamma_0) \delta_{pq}$, 
where $E_p$ is the energy eigenvalue of an eigenstate $|E_p \rangle$.
Then, from Eq.~(\ref{eq:Hnnn}) we see that
$H_{\rm nn}$ and $H_{\rm nnn}$ can be diagonalized simultaneously
by the basis of $|E_p \rangle$,
and the energy eigenvalue of the total Hamiltonian,
$H_{\rm nn}+H_{\rm nnn}$, is given by
\begin{align}
 E_p -\gamma_n \left( \frac{E_p}{-\gamma_0} \right)^2 + 3\gamma_n,
 \label{eq:hpert1}
\end{align}
for $|E_p \rangle$.
In Eq.~(\ref{eq:Hnnn}),
$H_{\rm nnn}$ contains the on-site potential part,
$3\gamma_n \sum_{i \in {\rm all}} c_i^\dagger c_i$.
This on-site potential can be ignored since it changes only
the origin of the energy band structure,
as shown by $3\gamma_n$ in Eq.~(\ref{eq:hpert1}).~\cite{wallace47}
This statement is correct for a periodic system without boundary,
but is not approved for a system with boundary.
It is because of that
the number of bonds of the edge sites is different from 
that of a bulk site and
the corresponding on-site potentials at the edge sites 
are different from those at the bulk sites.

To show this explicitly, 
we put an electron on the zigzag edge site
labeled as 2 in Fig.~\ref{fig:nnn}(c).
The electron is transfered to the nnn sites 
by ${\cal H}_{\rm nn}^2$.
For this time, however, 
the matrix element of the on-site potential term 
that we need to subtract from 
${\cal H}_{\rm nn}^2$ in order to get ${\cal H}_{\rm nnn}$
is $2$ because the number of bonds is 2 for the edge site.
It is different from $3$ for a non-edge (bulk) site. 
Thus, we obtain the formula for $H_{\rm nnn}$ as
\begin{align}
 H_{\rm nnn} 
 = -\gamma_n \sum_{i,j\in {\rm all}}
  c_i^\dagger \{ 
 [{\cal H}^2_{\rm nn}]_{ij} - g_i [I]_{ij} \}
 c_j,
 \label{eq:Hnnn_sub}
\end{align}
where $g_i$ is the number of bonds of $i$-th site.
Since we can shift the origin of the energy by $3\gamma_n$
without a loss of generality, 
Eq.~(\ref{eq:Hnnn_sub}) can be written as
\begin{align}
 H_{\rm nnn} 
 = -\gamma_n 
 \sum_{i,j\in {\rm all}}
  c_i^\dagger [{\cal H}^2_{\rm nn}]_{ij} c_j
 +\gamma_n \sum_{i\in {\rm all}}(g_i-3) \hat{n}_i,
 \label{eq:Hnnn_g}
\end{align}
where $\hat{n}_i=c_i^\dagger c_i$ 
is the number operator of $i$-th site.
The first term in Eq.~(\ref{eq:Hnnn_g})
(or the second term of Eq.~(\ref{eq:hpert1}))
shows that $H_{\rm nnn}$ breaks the particle-hole symmetry 
of $H_{\rm nn}$ because 
${\cal H}_{\rm nnn}$ contains the square of ${\cal H}_{\rm nn}$.
The second term in Eq.~(\ref{eq:Hnnn_g})
represents quantum well potentials at the edge sites 
with potential depth of $-\gamma_n$
because $g_i=2$ for a zigzag edge site.
The quantum well potential 
appears only at an edge site whose number of bonds 
is different from those of a bulk site.
Therefore, a surface state appearing near the boundary
is strongly affected by this potential.
For bulk states, 
only the first term of the right-hand side 
of Eq.~(\ref{eq:Hnnn_g}) is important.
In fact, if a system has no boundary 
(if a system is periodic), 
the second term of Eq.~(\ref{eq:Hnnn_g})
disappears and the nnn Hamiltonian is given only 
by the first term.
Therefore the first term can be considered as the bulk part 
and the second term is as the edge part of the nnn Hamiltonian,
i.e., 
$H_{\rm nnn} = H_{\rm nnn}^{\rm bulk} + H_{\rm nnn}^{\rm edge}$,
where
\begin{align}
 \begin{split}
  & H_{\rm nnn}^{\rm bulk} \equiv
  -\gamma_n \sum_{i,j\in {\rm all}}
  c_i^\dagger [{\cal H}^2_{\rm nn}]_{ij} c_j, \\
  & H_{\rm nnn}^{\rm edge} \equiv 
  \gamma_n \sum_{i\in {\rm all}}(g_i-3) \hat{n}_i.
 \end{split}
\end{align}

\begin{figure}[htbp]
 \begin{center}
  \includegraphics[scale=0.6]{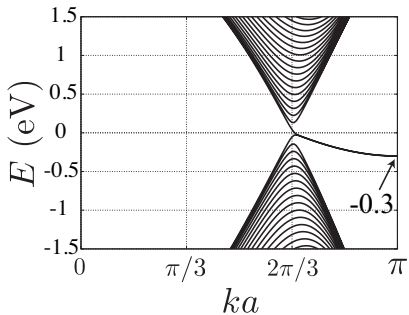}
 \end{center}
 \caption{The energy band structure of graphene system with the zigzag
 edge. This plot is obtained by diagonalizing $H_{\rm nn}+H_{\rm
 nnn}-3\gamma_n$ numerically. 
 We adopt $\gamma_0=3.0$ eV and $\gamma_n=0.3$ eV.
 The horizontal axis is a wavevector along the zigzag edge ($k$) 
 multiplied by the lattice constant ($a$).
 }
 \label{fig:band}
\end{figure}

The edge state is labeled by the wavevector along the zigzag edge,
$k$, as $| E(k) \rangle$.
Here $E(k)$ denotes the energy eigenvalue of $H_{\rm nn}$. 
Since $E(k)$ of the edge state
is very close to zero,~\cite{fujita96} 
the energy shift due to $H_{\rm nnn}^{\rm bulk}$, 
$-(\gamma_n/\gamma_0^2)E(k)^2$, is negligible.
Thus, the energy correction to the edge state
arises from $H_{\rm nnn}$ as
\begin{align}
 \Delta E(k) 
 = -\gamma_n \sum_{i \in {\rm edge}} 
 \langle E(k) | \hat{n}_i | E(k) \rangle.
 \label{eq:enedisp}
\end{align}
It is only the density at the edge sites that
determines $\Delta E(k)$.
The energy bandwidth ($W$) of the edge states  
can be calculated in the following way.
As $k$ approaches to the Fermi point,
i.e., $ka \to 2\pi/3$ or $4\pi/3$ ($a$ is lattice constant), 
the edge state changes into a bulk state
since the localization length 
$\xi(k) \to \infty$.~\cite{fujita96,sasaki05prb} 
Then $\sum_{i \in {\rm edge}} \langle E(k)| \hat{n}_i|E(k) \rangle$
is negligible and $\Delta E(2\pi/3a) = \Delta E(4\pi/3a)=0$.
On the other hand, 
the $k = \pi/a$ state is the most localized state
satisfying $\xi(\pi/a) = 0$.
For this state, we have
$\sum_{i \in {\rm edge}} \langle E(k)|\hat{n}_i |E(k)\rangle = 1$
and $\Delta E(\pi/a) = -\gamma_n$.
It shows that the $k = \pi/a$ state (the Fermi point) is located 
at the bottom (top) of the energy band
and $W=\gamma_n$ for the edge state.
This result of $W=\gamma_n$
reproduces the energy bandwidth that is numerically
calculated for the zigzag edge shown in
Fig.~\ref{fig:band}.~\cite{sasaki06apl} 
The quantum well potential of $H_{\rm nnn}^{\rm edge}$
lowers the edge state's energy if $\gamma_n$ is a positive value.
We adopt $\gamma_n \approx 0.3$ eV. 
This value is obtained by a first-principles calculation 
with the local density approximation.~\cite{porezag95}

\begin{figure}[htbp]
 \begin{center}
  \includegraphics[scale=0.5]{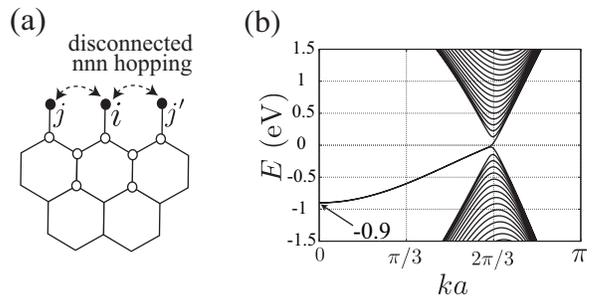}
 \end{center}
 \caption{
 (a) The lattice structure of the Klein edge. 
 The nnn hopping between nnn edge sites ($i$ and $j$ or $i$ and $j'$)
 is not represented by the double of the nn hopping.
 (b) The energy band structure of graphene system with the Klein edge. 
 } 
 \label{fig:klein}
\end{figure}

It should be mentioned that
Eqs.~(\ref{eq:Hnnn_sub}) and (\ref{eq:Hnnn_g}) 
include nnn hopping between sites which are connected
by two successive nn hopping processes.
Two successive nn hopping processes
do not include {\it disconnected} nnn hopping process.
The disconnected nnn process is relevant to the Klein edge
as shown in Fig.~\ref{fig:klein}(a).~\cite{klein94}
Because ${\cal H}_{\rm nn}^2$ can not transfer
an electron at the edge site ($i$) 
to the nnn edge site ($j$ or $j'$) in Fig.~\ref{fig:klein}(a), 
we have to add ${\cal H}_{\rm nnn}^{\rm dc}$,
which represents the nnn hopping between disconnected sites,
to the matrix of the right-hand side of Eq.~(\ref{eq:Hnnn_sub})
in order to get a complete ${\cal H}_{\rm nnn}$. 
Thus we have
\begin{align}
 H_{\rm nnn} 
 = -\gamma_n \sum_{i,j\in {\rm all}}
  c_i^\dagger \{ 
 [{\cal H}^2_{\rm nn}]_{ij} - g_i [I]_{ij}
 + [{\cal H}_{\rm nnn}^{\rm dc}]_{ij} \} c_j.
 \label{eq:Hnnn_c}
\end{align}
Since $[{\cal H}_{\rm nnn}^{\rm dc}]_{ij}$ is not zero only when
$i$ and $j$ are both the edge sites,
the disconnected nnn Hamiltonian is written as
\begin{align}
 H_{\rm nnn}^{\rm dc} 
 &\equiv -\gamma_n \sum_{i,j\in {\rm all}}
 c_i [{\cal H}_{\rm nnn}^{\rm dc}]_{ij} c_j \nonumber \\
 &= -\gamma_n \sum_{i,j \in {\rm edge}}^{{\rm nnn}} c_i^\dagger c_j.
\end{align}
$H_{\rm nnn}^{\rm dc}$ can be classified into 
the edge part of the nnn Hamiltonian since 
$H_{\rm nnn}^{\rm dc}$ is given by the
creation and annihilation operators at the edge sites.
If we represent the wavefunction 
using the density $n$ and phase $\theta$ as 
\begin{align}
 |\Psi \rangle =
 \sum_{i \in {\rm all}} \sqrt{n}_i e^{i\theta_i}
 c_i^\dagger|0\rangle,
\end{align}
then we have
\begin{align}
 \langle \Psi| H_{\rm nnn}^{\rm dc} |\Psi \rangle 
 = -\gamma_n \sum_{i,j \in {\rm edge}}^{\rm nnn} \sqrt{n_i n_j}
 e^{i(\theta_i - \theta_j)}.
 \label{eq:denpha}
\end{align}
This result shows that not only the density ($\sqrt{n_i n_j}$) 
but also the relative phase ($\theta_i - \theta_j$)
of the localized wavefunction is important for the energy shift.
This is contrasted to the fact 
that the quantum well potential couples only 
to the density of a quantum state.
If there is a lattice periodicity along the edge,
we can set $\theta_i=(ka)i$ and $n_j = n_i$. 
Then Eq.~(\ref{eq:denpha}) becomes
\begin{align}
 \langle \Psi| H_{\rm nnn}^{\rm dc} |\Psi \rangle 
 = -2 \gamma_n \cos (ka)
 \sum_{i \in {\rm edge}} n_i.
 \label{eq:h_nnn_k}
\end{align}

Since $g_i=1$ for the Klein edge sites,
the edge part of the nnn Hamiltonian is written as
\begin{align}
 H_{\rm nnn}^{\rm edge} 
 = - 2\gamma_n \sum_{i \in {\rm edge}} \hat{n}_i
 -\gamma_n \sum_{i,j \in {\rm edge}}^{\rm nnn} c_i^\dagger c_j.
\end{align}
Using Eq.~(\ref{eq:h_nnn_k}),
we get $\Delta E(k)$ 
($\equiv \langle E(k) | H_{\rm nnn}^{\rm edge} |E(k) \rangle$)
for the Klein edge as
\begin{align}
 \Delta E(k)
 = - 2\gamma_n (1+\cos(ka))
 \sum_{i \in {\rm edge}} n_i(k),
 \label{eq:H_klein}
\end{align}
where $n_i(k) \equiv \langle E(k)|\hat{n}_i|E(k) \rangle$.
The energy bandwidth for the Klein edge states
is calculated as follows.
Near the Klein edges, the edge states appear
for $0 \le k < 2\pi/3a$ and $4\pi/3a < k \le 2\pi/a$
(see Fig.~\ref{fig:klein}(b)).
It can be shown that 
the wavefunction of most localized state is given by $k=0$ state, 
and sum of the densities at the Klein edge sites is given by 
$\sum_{i \in {\rm edge}} n_i(0) = 3/4$.~\cite{sasaki05prb} 
Then, by putting $k=0$ into Eq.~(\ref{eq:H_klein}),
we have $\Delta E(0) = - 3\gamma_n$.
Thus, $W$ for the Klein edge is $3\gamma_n$ (=0.9eV).
This result also reproduces the energy bandwidth that is numerically
calculated for the Klein edge shown in Fig.~\ref{fig:klein}(b).
The half of $W$ comes from the quantum well potential 
and the rest half of $W$ is due to the disconnected nnn 
hopping process.

\begin{figure}[htbp]
 \begin{center}
  \includegraphics[scale=0.7]{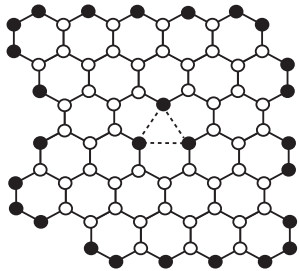}
 \end{center}
 \caption{
 A graphene system with a boundary and a lattice vacancy.
 $H_{\rm nnn}^{\rm edge}$ gives an on-site potential 
 energy shift of $-\gamma_n$ 
 at carbon atoms denoted by solid circles.
 There is a disconnected nnn Hamiltonian at the dotted lines around the
 lattice vacancy.
 } 
 \label{fig:vacancy}
\end{figure}

Here let us summarize the formula for the nnn Hamiltonian: 
$H_{\rm nnn}$ can be decomposed into bulk and edge parts as
$H_{\rm nnn} = H_{\rm nnn}^{\rm bulk} + H_{\rm nnn}^{\rm edge}$ 
with
\begin{align}
 \begin{split}
  & H_{\rm nnn}^{\rm bulk} \equiv
  -\gamma_n \sum_{i,j\in {\rm all}}
  c_i^\dagger [{\cal H}^2_{\rm nn}]_{ij} c_j, \\
  & H_{\rm nnn}^{\rm edge} \equiv 
  \gamma_n \sum_{i\in {\rm edge}} (g_i-g) \hat{n}_i
  -\gamma_n \sum_{i,j\in {\rm edge}}
  c_i^\dagger [{\cal H}^{\rm dc}_{\rm nnn}]_{ij} c_j.
 \end{split}
 \label{eq:Hdens}
\end{align}
$g$ is the number of bonds of a bulk site.
The first term of the right-hand side of $H_{\rm nnn}^{\rm edge}$
in Eq.~(\ref{eq:Hdens}) represents 
quantum well potentials at the edge sites. 
It is only the number of bonds at the edge site 
which determines the depth of the quantum well potential.
As a result, the quantum well potential appears 
at the edge sites, regardless of the edge shape.
For example, in the case of graphene, 
$g_i\ne g(=3)$ not only for the zigzag (or Klein) edge sites
but also for the armchair edge sites.
Thus, for a finite system of graphene shown in 
Fig.~\ref{fig:vacancy},
the quantum well potentials
of $-\gamma_n$ are denoted by the solid circles.
The quantum well potential at the armchair edge
may be not of importance as the zigzag edge 
because the edge state of a graphene is absent 
from the armchair edge.~\cite{fujita96,niimi05,kobayashi05} 
It is also interesting to note that
$g_i \ne g$ for the three sites around a lattice vacancy.
In addition to the quantum well potentials at the three sites,
the disconnected nnn hopping appears between them.

We remark on other effects 
that can modify the energy spectrum of the edge states.
First, it is naively expected that 
the orbital energy at an edge carbon atom 
is different from that at a bulk atom
when a functional group attaches to the edge atom.
The attachment of a functional group 
gives rise to an additional change of 
the energy bandwidth of the edge states.
However, this energy shift 
may be positive or negative value
depending on the type of a functional group.
This can be distinguished from the energy shift due to 
the nnn Hamiltonian because it is always a negative value.
Second, 
$W$ can be modified by the electron-electron or electron-phonon
interactions because they give rise to a self-energy correction 
to the edge states. 
A theoretical calculation of the self-energy for the edge states
is given in Refs.~\onlinecite{sasaki07local} 
and~\onlinecite{sasaki08jpsj}. 

We note that
Eq.~(\ref{eq:Hdens}) is not restricted to graphene systems
but is applicable to other two-dimensional systems
like the square lattice,
and three-dimensional systems.
The nnn Hamiltonian stabilizes surface states in a general system
through the quantum well potentials.
The energy dispersion relation of surface states
is observed below the Fermi level 
by high-resolution photo-emission studies
of the (111) surfaces of 
copper, silver and gold.~\cite{kevan87}
We speculate that the observed stability of the surface states 
is due to the edge part of the nnn Hamiltonian.

In conclusion,
$H_{\rm nnn}$ can be decomposed into the bulk and edge parts
as shown in Eq.~(\ref{eq:Hdens}).
If the energy spectrum of $H_{\rm nn}$ is symmetric 
with respect to $E=0$,
then $H_{\rm nnn}^{\rm bulk}$ breaks this symmetry.
If $H_{\rm nn}$ has a localized energy eigenstate near the edge of
a system, then $H_{\rm nnn}^{\rm edge}$ is relevant to shift the energy
eigenvalue through the quantum well potential and the disconnected nnn
edge Hamiltonian.
The quantum well potential couples only to the density,
whereas the disconnected nnn edge Hamiltonian 
couples to the phase of the localized wavefunction.
Although the Hamiltonian decomposition is proved for two-dimensional
graphene systems, it works for other systems like a one-dimensional
chain of atoms and a three dimensional lattice system as well.

\section*{Acknowledgment}

This work is financially supported by 
a Grant-in-Aid for Specially Promoted Research
(No.~20001006) from MEXT.

\bibliographystyle{apsrev}

\end{document}